\documentclass[3p,lefttitle,times]{elsarticle}

\journal{Applied Mathematical Modelling}

\usepackage{graphicx,xcolor}
\usepackage{hyperref}
\usepackage{amsmath,amssymb,cleveref}
\usepackage{textcomp} 

\usepackage{natbib}
\biboptions{sort&compress}

\definecolor{myred}{RGB}{160,0,0}
\definecolor{mygreen}{RGB}{0,160,0}
\definecolor{myblue}{RGB}{0,0,160}

\hypersetup{
    colorlinks = true,
    linkcolor = {myred},
    citecolor = {mygreen},
    urlcolor  = {myblue},
}

\newcommand{\RED}{} 

\begin{document}

\begin{frontmatter}

\title{Manipulating thermal fields with inhomogeneous heat spreaders}

\author{Eleanor R. Russell \corref{cor1}}
\ead{eleanor.russell@manchester.ac.uk}
\author{Rapha\"{e}l C. Assier}
\author{William J. Parnell}

\cortext[cor1]{Corresponding author}

\address{Department of Mathematics, The University of Manchester, Oxford Road, Manchester, M13 9PL, UK}

\begin{abstract}
We design a class of spatially inhomogeneous heat spreaders in the context of steady-state thermal conduction leading to spatially uniform thermal fields across a large convective surface. Each spreader has a funnel-shaped design, either in the form of a trapezoidal prism or truncated cone, and is forced by a thermal source at its base. We employ transformation-based techniques, commonly used to study metamaterials, to determine the required thermal conductivity for the spreaders. The obtained materials, although strongly anisotropic and inhomogeneous, are accurately approximated by assembling isotropic, homogeneous layers, rendering them realisable. An alternative approach is then considered for the conical and trapezoidal spreaders by dividing them into two or three isotropic, homogeneous components respectively. We refer to these simple configurations as neutral layers. All designs are validated numerically \RED{both with and without the effects of thermal contact resistance between interfaces}. Such novel designs pave the way for future materials that can manipulate and control the flow of heat, helping to solve traditional heat transfer problems such as controlling the temperature of an object and energy harvesting.
\end{abstract}


\end{frontmatter}

\section{Introduction}
Heat transfer studies primarily focus on temperature control and heat flux management. Traditional applications include energy harvesting and heating or cooling objects. In particular, designing thermal management devices for effective heat removal from electronics proves a significant challenge for engineers. In this work we propose a class of funnel-shaped heat spreaders \RED{which can be employed for heat sink applications}. Each spreader has the form of either a trapezoidal prism or truncated cone with a large convective surface at the top that is forced by a smaller thermal source at its base. Assuming all other external surfaces are perfectly insulated, we use mathematical modelling to optimise the design of each spreader such that, in a steady-state regime, thermal fields are spatially uniform across the larger convective surface, eliminating any temperature variation.

In general, anisotropic and inhomogeneous properties are required to guide the flow of heat through unconventional geometries in a spatially uniform manner. Here we demonstrate how to employ transformation-based techniques to determine the required anisotropic, inhomogeneous conductivity for the funnel-shaped spreaders. These methods, in principle, can be employed to achieve any desired thermal fields through any geometry. The prospect of realising specific physical fields with engineered materials, referred to as metamaterials, has led to progressively more research into transformation-based techniques over the past two decades. The first metamaterials were designed in order to control and manipulate electromagnetic wave fields. The designs for these metamaterials were based on transformation optics - a theory centred around the form invariance of Maxwell's equations after a general spatial transformation \cite{Pendry}. Following this, and due to the form invariance of various other governing equations, significant progress was also made with regard to the manipulation of other wave fields \cite{Banerjee_Intro_to_metamat} as well as thermal fields with transformation thermodynamics \cite{Li2021}. 

To date, transformation thermodynamics has assisted in tackling traditional heat transfer problems, with concepts such as thermal concentrators \cite{Guen2012,Jia2020,Ji2021,Narayana2012,Chen2015,Dede2018,Han2015,Hu2018}, lenses \cite{Han2015,Vemuri2014,Bandaru2015} and uniform heating devices \cite{Han2015,Liu2014,Han2018}. It has also enabled the development of new and exciting concepts such as thermal invisibility cloaks \cite{Guen2012,Jia2020,Narayana2012,Dede2018,Schittny2013,Han2014,Han2018,Hu2018,Zhu2021}, ground cloaks \cite{Yang2016,Hu2015,Qin2019}, camouflaging \cite{Han2014,Hu2021}, rotators \cite{Jia2020,Narayana2012,Dede2018,Guenneau2013,Zhu2021} \RED{and communication \cite{Hu2018,Hu2019}.} The most relevant of these concepts, in the context of this article, are: the heat plate designed by Liu et al.\ \cite{Liu2014}; the ground cloaks designed by Yang et al.\ \cite{Yang2016} and Hu et al.\ \cite{Hu2015}; \RED{and the thermal expander designed by Han et al.\ \cite{Han2018}}. In particular, we apply transformations that are analogous to those used in these papers to determine the required anisotropic, inhomogeneous properties of the spreaders. Once these properties have been obtained, we apply effective medium theory \RED{while considering the effects of thermal contact resistance \cite{Auriault1994,Hasselman1987}} to propose layered designs comprising isotropic, homogeneous materials that can be realised to approximate the required behaviour. This approach has been successfully applied, using combinations of both isotropic and anisotropic materials, to manufacture thermal metamaterials and validate their performance experimentally \cite{Narayana2012,Bandaru2015,Liu2014,Schittny2013,Yang2016}, \RED{however, the effects of contact resistance are not always considered. For example, in \cite{Liu2014}, the imperfection in the metamaterial fabrication could induce additional interfacial thermal resistance and lead to a reduced rise in temperature when experiments are compared to simulations. By incorporating interfacial effects into our models we are able to account for an additional temperature drop.}

\RED{We then move away from transformation theory and} consider an alternative approach where the spreader is divided into two or three isotropic, homogeneous components. These simple configurations, which we refer to as \textit{neutral layers}, are inspired by neutral inclusions in the context of thermal conductivity \cite{Benveniste1999,Benveniste2003} and elasticity \cite{Wang2012,Norris2020} \RED{in the sense that we exploit the solution to the diffusion equation for this specific configuration. This approach has been successfully applied to design a range of elliptical ground cloaks \cite{Qin2019} and an elliptical spreader \cite{Han2018}, however, the simple triangular geometry of our designs, which we evaluate with and without curvature, are yet to be considered.} The main difference between the neutral layer designs and the metamaterials obtained through transformation thermodynamics is that the neutral layers are tailored for a specific set of boundary conditions, whereas the metamaterials work for a range of boundary conditions. Whilst the metamaterials are more flexible in this sense, the simple neutral layer designs can be seen as a more practical approach. We validate all our designs numerically and, although still at the conceptual stage, such designs can help to pave the way for future materials that can manipulate and control the flow of heat.



\begin{figure*}[t!]
	\begin{center}
	\includegraphics[width=0.8\textwidth]{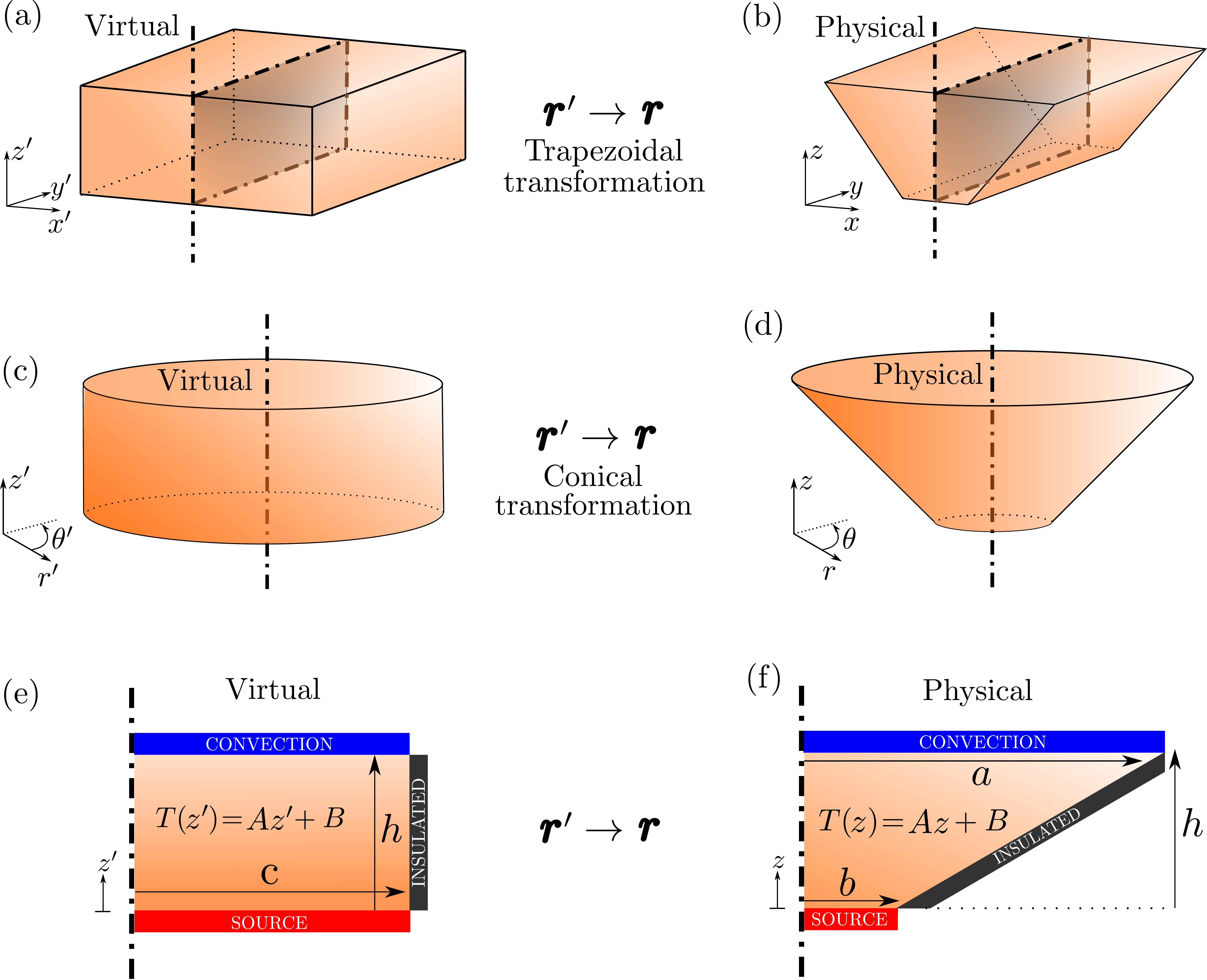}
	\caption{Transformation process for: (a)-(b) trapezoidal spreader; (c)-(d) conical spreader; (e)-(f) two-dimensional cross section of each spreader with imposed boundary conditions. The cross section is extended in the $y$-direction and reflected in the $yz$-plane to achieve the trapezoidal design or rotated about the $z$-axis to achieve the conical design.}
	\label{fig: transformation}
	\end{center}
\end{figure*}

\section{Transformation-based methods}
\subsection{Transformation theory} 
We focus on controlling steady-state thermal conduction which, in the absence of thermal sources or sinks, is governed by the steady-state heat diffusion equation
\begin{equation}
    \nabla' \cdot (\boldsymbol{k}' \nabla' T) = 0,
\label{eqn: heat eqn}
\end{equation}
where $\nabla'$ is the gradient, $\boldsymbol{k}'(\boldsymbol{r}')$ is the thermal conductivity tensor and $T(\boldsymbol{r}')$ is the temperature field where $\boldsymbol{r}'(\xi_i')$ is the position vector with coordinates $(\xi_1',\xi_2',\xi_3')$.

By following Banerjee's approach for transformation-based methods on electrical conductivity \cite{Banerjee_Intro_to_metamat}, it can be shown that the form of \eqref{eqn: heat eqn} is invariant after a general spatial transformation from a virtual space $\boldsymbol{r}'(\xi_i')$ to a physical space $\boldsymbol{r}(\xi_i)$ on the condition that the conductivity in the physical space, denoted by $\boldsymbol{k}$, satisfies
\begin{equation}
\boldsymbol{k} = \frac{\mathbf{F} \boldsymbol{k}' \mathbf{F}^T}{\det (\mathbf{F})},
\label{eqn: k relation}
\end{equation}
where $\mathbf{F} = \boldsymbol{r} \otimes \nabla'$ is the deformation gradient of the transformation $\boldsymbol{r}' \rightarrow \boldsymbol{r}$ where $\otimes$ is the tensor product. The result in \eqref{eqn: k relation} enables us to solve the inverse problem where, for a given geometry and set of boundary conditions, we can achieve specific thermal fields by engineering the conductivity. For example, here we focus on achieving spatially uniform thermal fields across the convective surface of a trapezoidal or conical spreader, illustrated by the top surfaces in Fig.~\ref{fig: transformation}(b) and (d) respectively. This is achieved by starting with the virtual domains illustrated in Fig.~\ref{fig: transformation}(a) and (c), both with an isotropic conductivity denoted by a constant scalar $k'$, and applying one of two transformations: a linear stretch affecting each point in the virtual space, or a modified ground cloak transformation that only affects half of the virtual space. These transformations have been chosen to preserve the form of the virtual temperature field which is linear with respect to $z$. As a result, the thermal fields in the spreaders only exhibit $z$-dependence. This leads to fields that are spatially uniform with respect to the other two coordinates, eliminating temperature variation across the convective surface as desired.

We first consider a heat spreader in the form of a trapezoidal prism. The full transformation process for this spreader is illustrated in Fig.~\ref{fig: transformation}(a)-(b) where we assume that the boundary conditions across the base and convective surface are homogeneous, and all side surfaces are perfectly insulated. With this assumption, we can focus on the two-dimensional design obtained by reflecting the transformation in Fig.~\ref{fig: transformation}(e)-(f) in the $z$-axis.

\subsection{Trapezoidal design}

\subsubsection{Linear stretch} \label{sec: linear Cart method}
For the following linear stretch mapping we set $c=b$ in Fig.~\ref{fig: transformation}(e). In other words, the base of the virtual and physical domain have the same length. We then transform the virtual domain by performing a linear stretch in the $x$-coordinate. The relevant mapping $\boldsymbol{r}'(x',z') \rightarrow \boldsymbol{r}(x,z)$ is given by
\begin{equation}
\label{eqn: linear map Cart}
    x = \left[ \dfrac{(a-b)}{bh}z' + 1 \right]x', \qquad  z = z',
\end{equation}
where $a,\ b$ and $h$ are shown in Fig.~\ref{fig: transformation}(f). The deformation gradient of a mapping $\boldsymbol{r}'(x',z') \rightarrow \boldsymbol{r}(x,z)$ is given by

\begin{equation} \label{eqn: F Cart}
\mathbf{F}= \left(x \textbf{e}_x +  z \textbf{e}_z\right) \otimes \left(\textbf{e}_{x'} \dfrac{\partial}{\partial x'} + \textbf{e}_{z'} \dfrac{\partial}{\partial z'}\right) = \left[ \begin{matrix}
\ \dfrac{\partial x}{\partial x'} & \dfrac{\partial x}{\partial z'} \ \\ \\
\ \dfrac{\partial z}{\partial x'} & \dfrac{\partial z}{\partial z'} \
\end{matrix}\right].
\end{equation}
Therefore, from \eqref{eqn: k relation}, the transformed conductivity of the spreader must satisfy

\begin{equation}
\boldsymbol{k} =  \dfrac{k'}{\beta_1} \left[ \begin{matrix}
\ \beta_1 & \beta_2 \ \\ \\
\ 0 & 1 \
\end{matrix}\right]\left[ \begin{matrix}
\ \beta_1 & 0 \ \\ \\
\ \beta_2 & 1 \
\end{matrix}\right] = \dfrac{k'}{\beta_1} \left[ \begin{matrix}
    \ \beta_1^2 + \beta_2^2 & \quad \beta_2 \ \\ \\
\beta_2 & \quad 1 \
    \end{matrix}\right],
    \label{eqn: k Cart linear}
\end{equation}
where
\begin{equation}
\beta_1(z) = \dfrac{\partial x}{\partial x'} = \dfrac{(a-b)z+ bh}{bh} \qquad \text{and} \qquad 
\beta_2(x,z) = \dfrac{\partial x}{\partial z'} = \dfrac{(a-b)x}{(a-b)z + bh},
\label{eqn: beta 1 2}
\end{equation}
once written in terms of the physical coordinates.

In Fig.~\ref{fig: iso vs metamat Cart}(a) we show a finite element simulation\footnote{All finite element simulations are performed with COMSOL Multiphysics\textregistered Version 5.5.} for the resulting thermal fields through an isotropic, homogeneous spreader with conductivity $k=80$W/(mK). Referring to Fig.~\ref{fig: transformation}(f), we set $a=5b=2h=10$cm and impose a constant temperature across the base, given by $T(z=0)=80^{\circ}$C, and a convective boundary condition across the top surface where the surrounding air has a heat transfer coefficient of $h_c=15$W/(m$^2$K) and temperature $T_0=20^{\circ}$C. Unless otherwise stated, each simulation in this article is subjected to these boundary conditions. The white lines in Fig.~\ref{fig: iso vs metamat Cart} represent isotherms. We see that the temperature is not uniform across the top surface for the isotropic case. In comparison, Fig.~\ref{fig: iso vs metamat Cart}(b) shows a simulation for a spreader with the same geometry and boundary conditions as Fig.~\ref{fig: iso vs metamat Cart}(a), but with an anisotropic, inhomogeneous conductivity that satisfies \eqref{eqn: k Cart linear} when $k'=80$W/(mK). The thermal fields are uniform with respect to $x$ for this transformed case, eliminating any temperature variation across the top surface, as desired.

\subsubsection{Modified ground cloak}

\begin{figure}[t!]
	\begin{center}
	\includegraphics[width=0.95\textwidth]{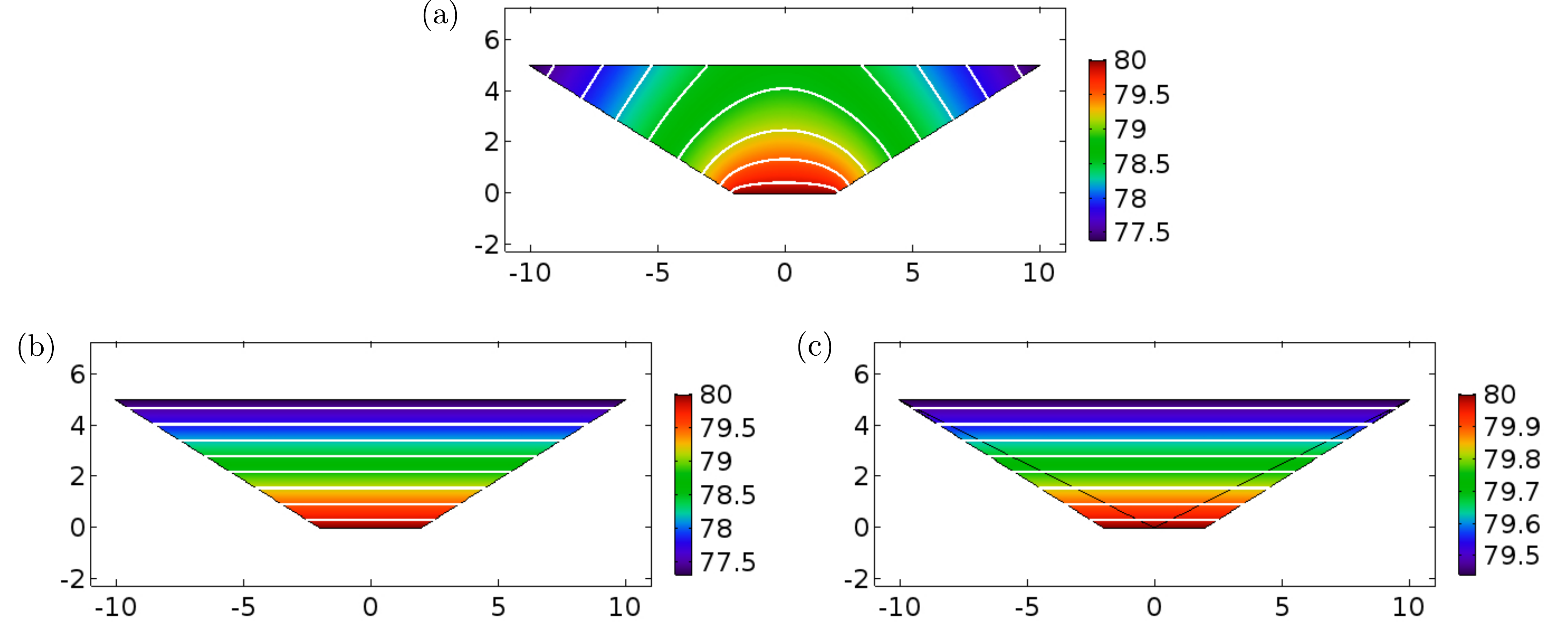}
	\caption{Simulations comparing trapezoidal heat spreaders with: (a) isotropic conductivity; (b) conductivity satisfying \eqref{eqn: k Cart linear}; (c) top component with isotropic conductivity and two layers beneath with conductivity satisfying \eqref{eqn: k con linear}. White lines represent isotherms.}
	\label{fig: iso vs metamat Cart}
	\end{center}
\end{figure}

For the following mapping we set $c=a$ in Fig.~\ref{fig: transformation}(e). In other words, the base of the virtual domain and the top surface of the physical domain have the same length. Traditionally, a ground cloak transformation compresses the virtual space away from the $z$-axis, creating a protected region around which the heat flux is guided. We employ this transformation to compress the virtual space towards the $z$-axis. \RED{By modifying the transformation in this way we have been able to utilise this mapping in a new heat-spreading context.} The relevant mapping $\boldsymbol{r}'(x',z')\rightarrow \boldsymbol{r}(x,z)$ only transforms points of the virtual space that satisfy $z' < h|x'|/a$ and is given by
\begin{equation}
\label{eqn: GC map}
x =  \dfrac{b}{a}x' + \textnormal{sgn}(x')\dfrac{a-b}{h}z',  \qquad z=z',
\end{equation}
where $a,\ b$ and $h$ are shown in Fig.~\ref{fig: transformation}(f). From \eqref{eqn: k relation} and \eqref{eqn: F Cart}, the conductivity in the transformed regions of the spreader must satisfy
\begin{equation}
\boldsymbol{k}  = \dfrac{k'}{\beta_3} \left[ \begin{matrix}
\ \beta_3 & \textnormal{sgn}(x)\beta_4 \ \\ \\
\ 0 & 1 \
\end{matrix}\right]\left[ \begin{matrix}
\ \beta_3 & 0 \ \\ \\
\ \textnormal{sgn}(x)\beta_4 & 1 \
\end{matrix}\right] = \dfrac{k'}{\beta_3} \left[ \begin{matrix}
    \ \beta_3^2 + \beta_4^2 & \quad \textnormal{sgn}(x)\beta_4 \ \\ \\
\textnormal{sgn}(x)\beta_4 & \quad 1 \
    \end{matrix}\right],
    \label{eqn: k Cart GC}
\end{equation}
where
\begin{equation}
\beta_3 = \dfrac{b}{a} \quad \text{and} \quad \beta_4 = \dfrac{a-b}{h}.
\end{equation}
In contrast to \eqref{eqn: k Cart linear}, although the transformed conductivity in \eqref{eqn: k Cart GC} is anisotropic, it is now homogeneous. Furthermore, any points of the virtual space that satisfy $z' \geq h|x'|/a$ are unaffected by the transformation and so the conductivity remains isotropic in this region with conductivity $k=k'$. For example, Fig.~\ref{fig: iso vs metamat Cart}(c) shows a simulation for a spreader with the same geometry and boundary conditions as the isotropic spreader in Fig.~\ref{fig: iso vs metamat Cart}(a), but this spreader is composed of three components: an isotropic component with conductivity  $k=k'=80$W/(mK) that lies on top, and two metamaterial layers that lie beneath with conductivity satisfying \eqref{eqn: k Cart GC}. We see that the thermal fields are spatially uniform with respect to $x$, eliminating temperature variation across the convective surface, as desired.

Next we consider a heat spreader in the form of a truncated cone. In order to simplify the realisation process for this design, we only consider the linear stretch transformation. The full transformation process for this spreader is illustrated in Fig.~\ref{fig: transformation}(c)-(d) where we assume that the boundary conditions across the base and convective surface are axisymmetric. With this assumption, this process is equivalent to rotating the transformation in Fig.~\ref{fig: transformation}(e)-(f) about the $z$-axis.

\subsection{Conical design}

\subsubsection{Linear stretch} As with the previous linear stretch mapping in Section~\ref{sec: linear Cart method}, we set $c=b$ in Fig.~\ref{fig: transformation}(e). Working in cylindrical coordinates, we then transform the virtual domain by performing a linear stretch in the $r$-coordinate. This transformation is equivalent to the linear stretch applied by Liu et al. to design a plate heater in a transient regime \cite{Liu2014}. The relevant mapping $\boldsymbol{r}'(r',\theta ',z') \rightarrow \boldsymbol{r}(r,\theta,z)$ is given by
\begin{equation}
\label{eqn: linear map con}
    r = \left[ \dfrac{(a-b)}{bh}z' + 1 \right] r', \qquad \theta = \theta', \qquad  z = z',
\end{equation}
where $a,\ b$ and $h$ are shown in Fig.~\ref{fig: transformation}(e)-(f). The deformation gradient, $\mathbf{F} = \boldsymbol{r} \otimes \nabla'$, of a mapping $\boldsymbol{r}'(r',\theta',z') \rightarrow \boldsymbol{r}(r,\theta,z)$, where $r(r',z'),\ \theta=\theta'$ and $z=z'$, is given by

\begin{equation}
\label{eqn: F cyl 1}
\begin{split}
\mathbf{F} &=  \left( r \textbf{e}_r + z \textbf{e}_z \right) \otimes \left(\textbf{e}_{r'} \dfrac{\partial}{\partial r'} + \textbf{e}_{\theta '} \dfrac{1}{r'}\dfrac{\partial}{\partial \theta '} + \textbf{e}_{z'} \dfrac{\partial}{\partial z'}\right) \\
& = \dfrac{\partial r}{\partial r'}\textbf{e}_{r} \otimes \textbf{e}_{r'} + \dfrac{r}{r'}\textbf{e}_{\theta} \otimes \textbf{e}_{\theta'} + \dfrac{\partial r}{\partial z'}\textbf{e}_{r} \otimes \textbf{e}_{z'} + \textbf{e}_{z} \otimes  \textbf{e}_{z'} = \left[ \begin{matrix}
\dfrac{\partial r}{\partial r'} & 0 & \dfrac{\partial r}{\partial z'}\\
0 & \dfrac{r}{r'} & 0 \\
0 & 0 & 1
\end{matrix}\right],
\end{split}
\end{equation}
in the basis $\{ \mathbf{e}_r,\ \mathbf{e}_\theta,\ \mathbf{e}_z \}$. Therefore, from \eqref{eqn: k relation}, the transformed conductivity must satisfy
\begin{equation}
\boldsymbol{k} =  \dfrac{k'}{\beta_1^2} \left[ \begin{matrix}
\beta_1 & 0 & \beta_5\\
0 & \beta_1 & 0 \\
0 & 0 & 1
\end{matrix}\right]\left[ \begin{matrix}
\beta_1 & 0 & 0 \\
0 & \beta_1 & 0 \\
\beta_5 & 0 & 1
\end{matrix}\right] = \dfrac{k'}{\beta_1^2} \left[ \begin{matrix}
    \beta_1^2 + \beta_5^2 & \ 0 \ & \quad \beta_5 \  \\
0 & \ \beta_1^2 \ & \quad 0 \  \\
\beta_5 & \ 0 \ & \quad 1 \
    \end{matrix}\right],
    \label{eqn: k con linear}
\end{equation}
where $\beta_1(z)$ is given in \eqref{eqn: beta 1 2} and
\begin{equation}
\beta_5(r,z) = \dfrac{\partial r}{\partial z'} = \dfrac{(a-b)r}{(a-b)z + bh},
\label{eqn: beta 3}
\end{equation}
once written in terms of the physical coordinates.

\begin{figure}[t!]
	\begin{center}
	\includegraphics[width=0.8\textwidth]{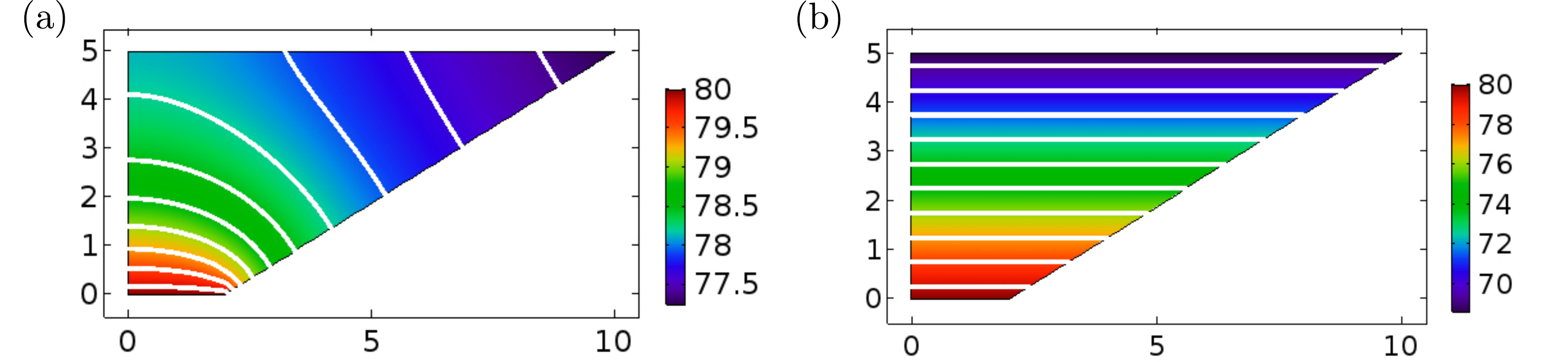}
	\caption{Simulations comparing conical heat spreaders with: (a) isotropic conductivity; (b) conductivity satisfying  \eqref{eqn: k con linear}. White lines represent isotherms.}
	\label{fig: iso vs metamat con}
	\end{center}
\end{figure}

Fig.~\ref{fig: iso vs metamat con}(a) shows a simulation for the cross section of an isotropic, homogeneous conical spreader with conductivity $k=80$W/(mK). Referring to  Fig.~\ref{fig: transformation}(f), we set $a=5b=2h=10$cm and impose the same boundary conditions as in Fig.~\ref{fig: iso vs metamat Cart}. In comparison, Fig.~\ref{fig: iso vs metamat con}(b) shows a simulation for a spreader with the same geometry and boundary conditions as Fig.~\ref{fig: iso vs metamat con}(a) but with an anisotropic, inhomogeneous conductivity that satisfies   \eqref{eqn: k con linear} when $k'=80$W/(mK). We see that the thermal fields are uniform with respect to $r$ for the transformed case in Fig.~\ref{fig: iso vs metamat con}(b), eliminating any temperature variation across the top surface, as desired. In what follows, we apply effective medium theory \RED{with and without the effects of thermal contact resistance} to approximate the anisotropic behaviour in \eqref{eqn: k Cart linear}, \eqref{eqn: k Cart GC} and \eqref{eqn: k con linear} with isotropic, homogeneous layers.

\section{Realisation with layered materials}
Anisotropic behaviour can be approximated with layered structures where the thickness and material properties of each layer are determined through effective medium theory. \RED{Furthermore, interfaces may play a major role in modulating heat transfer in thermal metamaterials \cite{Bandaru2015}, and so we incorporate thermal contact resistance into our effective medium theory model by assuming that a temperature jump occurs across each interface \cite{Auriault1994,Hasselman1987}. In particular, the temperature jump across an interface that lies perpendicular to the $z$-axis, denoted by $[\![ T ]\!]_{z_I}$ when $z=z_I$, can be expressed as
\begin{equation}
    [\![ T ]\!]_{z_I} = T(z_I^-) - T(z_I^+) = R_c \left( -k \dfrac{\partial T}{\partial z}\right) \biggr\rvert _{z=z_I^+},
    \label{eqn: temp jump}
\end{equation}
where $R_c$ is the thermal resistance of the interface. Although theoretical methods have been developed to predict $R_c$, the most reliable results are achieved experimentally and are typically in the range of 0.1-4$\times10^{-4}$m$^2$K/W for metallic interfaces at contact pressures of 10,000~kN/m$^2$ \cite{incropera1996}.

Here, we consider bilayered designs composed of two isotropic, homogeneous materials with conductivities denoted by the scalars $k_A$ and $k_B$. By assuming continuity of flux and a temperature drop, given by \eqref{eqn: temp jump}, across each interface,} the effective conductivities parallel and perpendicular to the layers, denoted by $k_\parallel$ and $k_\perp$ respectively, are given by
\begin{equation}
    k_\parallel = \phi k_A + (1-\phi)k_B \qquad \text{and} \qquad \dfrac{1}{k_\perp} = \dfrac{\phi}{k_A} + \dfrac{1-\phi}{k_B} \RED{+ \dfrac{2R_c}{L}},
    \label{eqn: eff k}
\end{equation}
where $\phi$ is the filling fraction of the material with conductivity $k_A$, \RED{$L$ is the total thickness of the bilayer and $2R_c$ accounts for both the resistance between the two layers and the resistance between two consecutive bilayers \cite{Auriault1994,Hasselman1987}. Note that the effects of contact resistance do not affect $k_\parallel$ \cite{Auriault1994}.} Assuming that $\phi$ is the same for $k_\parallel$ and $k_\perp$, we can solve the equations in \eqref{eqn: eff k} simultaneously to obtain
\begin{equation}
    k_B = \dfrac{k_\perp(k_A-k_\parallel)}{k_A-k_\perp \RED{- \frac{2R_c}{L}k_A k_\perp}}.
    \label{eqn: kB}
\end{equation}
Therefore, once we obtain the required effective conductivities from transformation thermodynamics, for a choice of $k_A > k_\parallel$, we can determine the necessary value for $k_B$ from \eqref{eqn: kB}. Then we rearrange one of the equations in \eqref{eqn: eff k} to find $\phi$, for example,
\begin{equation}
    \phi = \dfrac{k_\parallel-k_B}{k_A-k_B}.
    \label{eqn: phi}
\end{equation}
\RED{Note that $L$ is restricted when incorporating $R_c$ in this way since we require $k_\parallel > k_B > 0$. In particular, by substituting $k_\parallel > k_b$ into \eqref{eqn: kB}, we obtain a lower bound on $L$ for the work in this article, given by
\begin{equation}
    L > 2R_c \dfrac{k_\parallel k_\perp}{k_\parallel - k_\perp},
    \label{eqn: L bound}
\end{equation}
where $k_\parallel$ and $k_\perp$ are obtained through transformation theory.}

\subsection{Trapezoidal design}
\subsubsection{Linear stretch}\label{sec: linear stretch approx}
We approximate the anisotropic, inhomogeneous conductivity in \eqref{eqn: k Cart linear} by dividing the spreader into $n$ sub-layers of equal height \RED{$L$ where $L=h/n$}. We let $\boldsymbol{k}_i$ denote the required anisotropic conductivity in layer $i$ (where $i=1,...,n$ with $i=1$ referring to the base layer) and approximate $\boldsymbol{k}_i$ by substituting
\begin{equation}
    z = \dfrac{h(2i-1)}{2n},
    \label{eqn: zi}
\end{equation}
into   \eqref{eqn: k Cart linear}. In other words, we set $z$ to be the average $z$-value in each layer. Therefore, this approximation is piece-wise constant in $z$ in the sense that $z$-dependence is removed from each sub-layer without removing the $z$-dependence from the spreader as a whole. To simplify further we remove $x$-dependence in each sub-layer by setting $\beta_2=0$. Another way of thinking about this is to divide the virtual domain into $n$ sub-layers and stretch each layer individually where the stretch applied to layer $i$ is obtained by substituting   \eqref{eqn: zi} directly into the transformation in   \eqref{eqn: linear map Cart}. In other words, in layer $i$ we apply the transformation
\begin{equation}
x = \left(\dfrac{(a-b)(2i-1) + 2nb}{2nb} \right)x', \qquad z=z'.
\end{equation}
Approximating the required conductivity tensor in this way leads to
\begin{equation}
    \boldsymbol{k}_i = k' \left[ \begin{matrix}
    \ \gamma_i & 0 \ \\
    0 & \gamma_i^{-1} \
    \end{matrix}\right]
    \qquad \text{where} \qquad \gamma_i = \dfrac{(a-b)(2i-1) + 2nb}{2nb}.
    \label{eqn: k linear Cart approx}
\end{equation}

\begin{figure*}[t!]
	\begin{center}
	\includegraphics[width=0.89\textwidth]{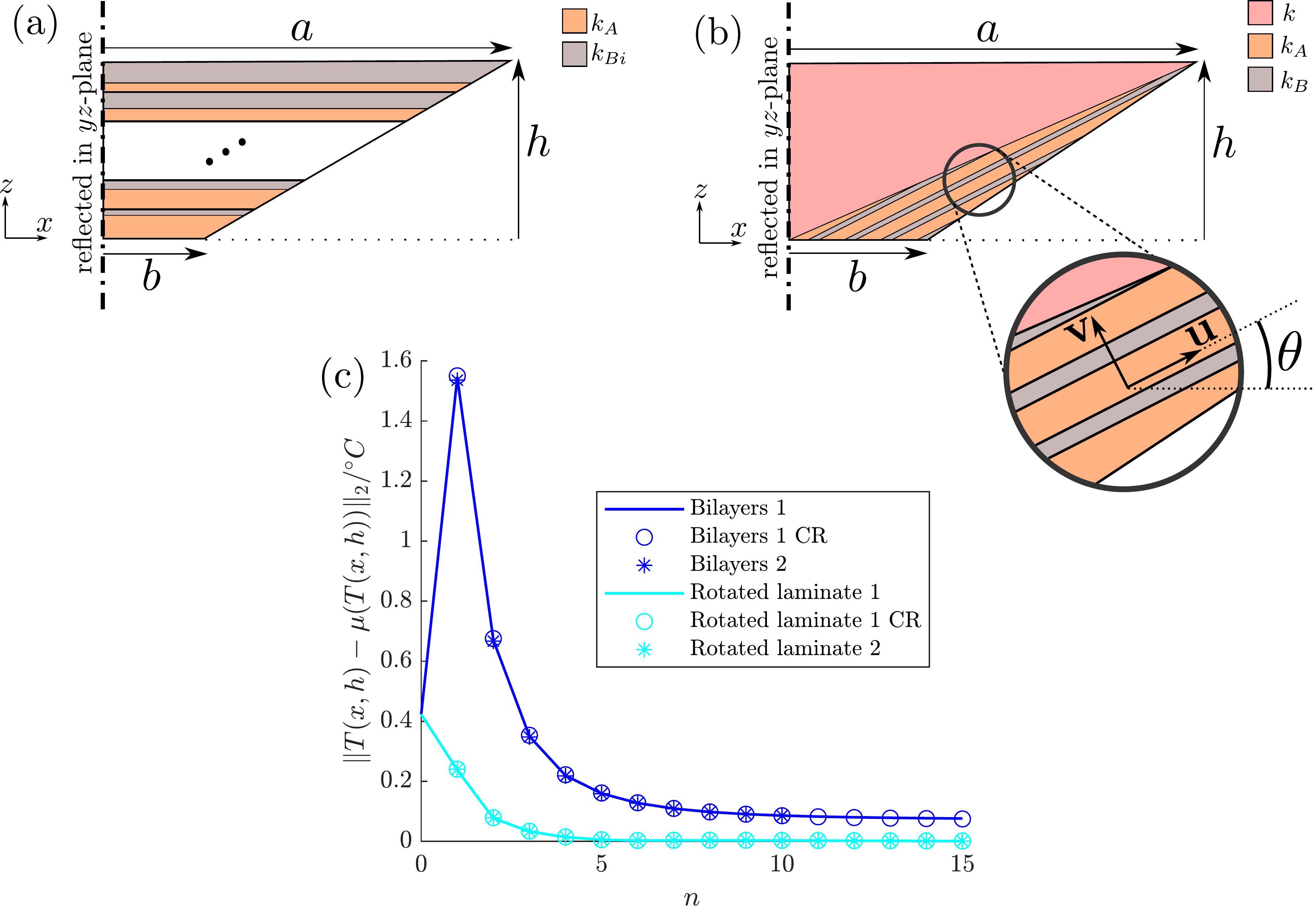}
	\caption{Trapezoidal heat spreader design using (a) bilayers and (b) a laminate rotated through some angle, $\theta$. (c) Simulations comparing the temperature variation across the top surface of each design using $n$ bilayers \RED{for three cases: a case designed for perfect contact; the same perfect contact design, but with thermal contact resistance between each layer; and a second design where $R_c$ is incorporated into $k_{Bi}$.}}
	\label{fig: 2D design}
	\end{center}
\end{figure*}

The anisotropic behaviour in \eqref{eqn: k linear Cart approx} is homogeneous and can therefore be approximated with a bilayered material using \eqref{eqn: eff k}. This design is illustrated in Fig.~\ref{fig: 2D design}(a) where the layers lie parallel to the $x$-axis. Each bilayer is composed of two isotropic, homogeneous layers. The conductivity of the bottom layer, denoted by $k_A$, is fixed throughout the design whereas the conductivity of the top layer, denoted by $k_{Bi}$ for layer {i}, varies from bilayer to bilayer. In particular, $k_{Bi}$ is calculated by substituting $k_A,\ k_\parallel=\gamma_i k'$ and $k_\perp=k'/\gamma_i$ into \eqref{eqn: kB}. The necessary filling fraction in each bilayer is then calculated from \eqref{eqn: phi}. For a given configuration we then run simulations for a range of $n$ bilayers and calculate the temperature variation across the top surface each time with respect to the L2-norm, which is defined as
\begin{equation*}
    \| \ T(x,h) -  \mu\left(T(x,h)\right) \ \|_2 = \sqrt{\mu\left(\left(T(x,h)-\mu\left(T(x,h)\right)\right)^2\right)}
\end{equation*}
for the trapezoidal case where $\mu$ is the average operator. 

\begin{figure*}[t!]
	\begin{center}
	\includegraphics[width=0.87\textwidth]{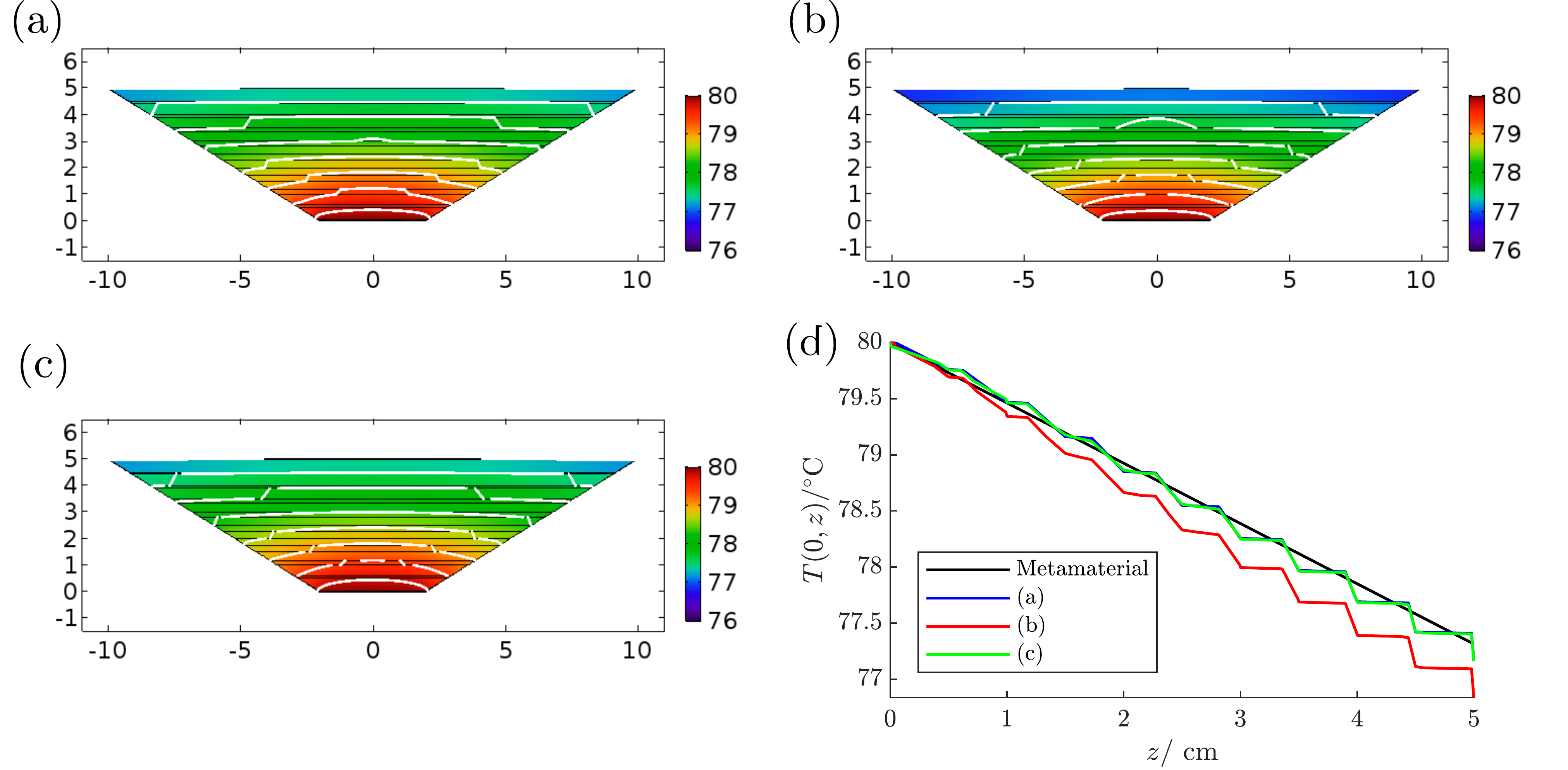}
	\caption{\RED{Simulations of a trapezoidal spreader designed using 10 bilayers with: (a) Perfect contact; (b) same design as for perfect contact, but with thermal contact resistance between each layer; (c) thermal contact resistance incorporated into $k_{Bi}$. (d) Comparison of the temperature drop through the centre of each case and the metamaterial in Fig.~\ref{fig: iso vs metamat Cart}(b).}}
	\label{fig: 2D results a}
	\end{center}
\end{figure*}

\RED{As a proof of concept, Fig.~\ref{fig: 2D design}(c) (blue) shows this process for three cases: a case designed for perfect contact, where $R_c=0\ $m$^2$K/W; the same perfect contact design, but with thermal contact resistance between each layer where $R_c=0.1 \times 10^{-4}$m$^2$K/W; and a second design where $R_c=0.1 \times 10^{-4}$m$^2$K/W is incorporated into $k_{Bi}$. Note that, when $n>10$, \eqref{eqn: L bound} is not satisfied for the second design and so there are no results for these cases.} For each case here we set $a=5b=2h=10$cm$,\ k'=80$W/(mK)$,\ k_A=5k'$ and impose the same boundary conditions as in Fig.~\ref{fig: iso vs metamat Cart}. \RED{We see that the effects of thermal contact resistance are negligible in terms of the temperature variation, however, it can lead to a reduction in temperature when compared to the perfect contact simulation. For example, from Fig.~\ref{fig: 2D design}(c), a suitable choice of $n$ is $n=10$ where a temperature variation of approximately $0.076^{\circ}$C is achieved for each case. Fig.~\ref{fig: 2D results a}(a)-(c) shows simulations for the three cases when 10 bilayers are used. The parameters for each bilayer in Fig.~\ref{fig: 2D results a}(a)-(b) and Fig.~\ref{fig: 2D results a}(c) are provided in columns 4-5 and 6-7 of Table~\ref{tab: 2D bilayers} respectively. Note that $k_A$ is omitted from Table~\ref{tab: 2D bilayers} as it is the same in each bilayer, namely $k_A=5k'=400$W/(mK). It is clear by comparing columns 4-5 and 6-7 that contact resistance has the largest effect on the bilayers at the base of the design, i.e.~the layers where more highly conducting materials are used.

Fig.~\ref{fig: 2D results a}(d) compares the temperature drop through the centre of each case against the metamaterial in Fig.~\ref{fig: iso vs metamat Cart}(b). We see that, when the effects of contact resistance are introduced into the perfect contact design, a larger temperature drop occurs, however, we have been able to account for this temperature drop by incorporating the effects of contact resistance into our model. As a result, the design is a better approximation of the metamaterial.}

\begin{table}[h!]
	\caption{Parameters for the bilayer simulations in Fig.~\ref{fig: 2D design}.}
	\centering
	\begin{tabular}{c c c c c c c}
	\hline
Bilayer & $k_\parallel \ /(W/(mK))$ &  $k_\perp \ /(W/(mK))$  &  $k_{Bi} \ /(W/(mK))$  & \hspace{.75cm}$\phi$\hspace{.75cm} &  \RED{$k_{Bi}^{CR} \ /(W/(mK))$}  & \hspace{.75cm}\RED{$\phi^{CR}$}\hspace{.75cm}\\
		 \hline
1  &  96  & 66.67 &	60.80 &	0.10 & \RED{89.41} & \RED{0.02}\\

2  &  128 &	50.00 &	38.86 &	0.25 & \RED{50.37} & \RED{0.22} \\

3  &  160 & 40.00 &	26.67 &	0.36 & \RED{32.43} & \RED{0.35} \\

4  &  192 &	33.33 &	18.91 &	0.45 & \RED{22.13} & \RED{0.45} \\

5  &  224 &	28.57 &	13.54 &	0.55 & \RED{15.44} & \RED{0.54} \\

6  &  256 &	25.00 &	9.60  &	0.63 & \RED{10.75} & \RED{0.63} \\

7  &  288 &	22.22 &	6.59  &	0.72 & \RED{7.27} & \RED{0.71} \\

8  &  320 &	20.00 &	4.21  &	0.80 & \RED{4.60} & \RED{0.80} \\

9  &  352 &	18.18 &	2.29  &	0.88 & \RED{2.47} & \RED{0.88} \\

10 &  384 &	16.67 &	0.70  &	0.96 & \RED{0.75} & \RED{0.96} \\
\hline
\end{tabular}
	\label{tab: 2D bilayers}
\end{table}

\subsubsection{Modified ground cloak}
The conductivity in \eqref{eqn: k Cart GC} is homogeneous. This somewhat simplifies the realisation process in the sense that $k_A,\ k_B$ and $\phi$ are fixed throughout the transformed region. Furthermore, the anisotropy in \eqref{eqn: k Cart GC} can be achieved with a periodic, laminated design that is rotated through some angle, $\theta$, as illustrated in Fig.~\ref{fig: 2D design}(b). The axes $\mathbf{u}$ and $\mathbf{v}$ in this design are referred to as the principal axes of the system. When the conductivity tensor is aligned with its principal axes it can be written in the form $\boldsymbol{k}=\textnormal{diag}(k_\parallel,k_\perp)$ where $k_\parallel$ and $k_\perp$ are referred to as the principle conductivities. The process of determining $k_\parallel,\ k_\perp$ and $\theta$ is described in \ref{app: princ sys}. Once obtained, we substitute $k_\parallel, \  k_\perp$ and our choice of $k_A$ into \eqref{eqn: kB} to obtain $k_B$. Then $\phi$ is determined by substituting $k_\parallel,\ k_A$ and $k_B$ into \eqref{eqn: phi}. For a given configuration we then run simulations for a range of $n$ layers, where $n$ now refers to how many times the layers are repeated, and calculate the temperature variation across the top surface with respect to the L2-norm. 

This process is shown \RED{for three cases} in Fig.~\ref{fig: 2D design}(c) (red): \RED{a case designed for perfect contact; the same perfect contact design, but with thermal contact resistance between each layer where $R_c=0.1 \times 10^{-4}$m$^2$K/W; and a second design where contact resistance is incorporated into $k_{Bi}$.} For each case we set $a=5b=2h=10$cm and  $k'=80$W/(mK) such that $k_\parallel=1436$W/(mK), $k_\perp=4.5$W/(mK) and $\theta \approx 35^\circ$. Furthermore, for each case we set $k_A=25k'$ and impose the same boundary conditions as in Fig.~\ref{fig: iso vs metamat Cart}. From Fig.~\ref{fig: 2D design}(c) we can choose a suitable choice for $n$. For example, a choice of $n=5$ achieves a temperature variation of approximately $0.0051^{\circ}$C for each case. \RED{Note that, when $n=5$, for the case of perfect contact, we obtain $k_B=1.28$W/(mK) and $\phi = 0.7$, whereas, for the case where $R_c=0.1 \times 10^{-4}$m$^2$K/W is incorporated into the design, we obtain $k_B^{CR}=1.31$W/(mK) and $\phi^{CR} = 0.7$. Therefore, the effects of contact resistance are negligible for this case. This is to be expected since the direction of the heat flux is no longer being perpendicular to the interfaces due to the laminate being rotated.}

\begin{figure*}[t!]
	\begin{center}
	\includegraphics[width=0.87\textwidth]{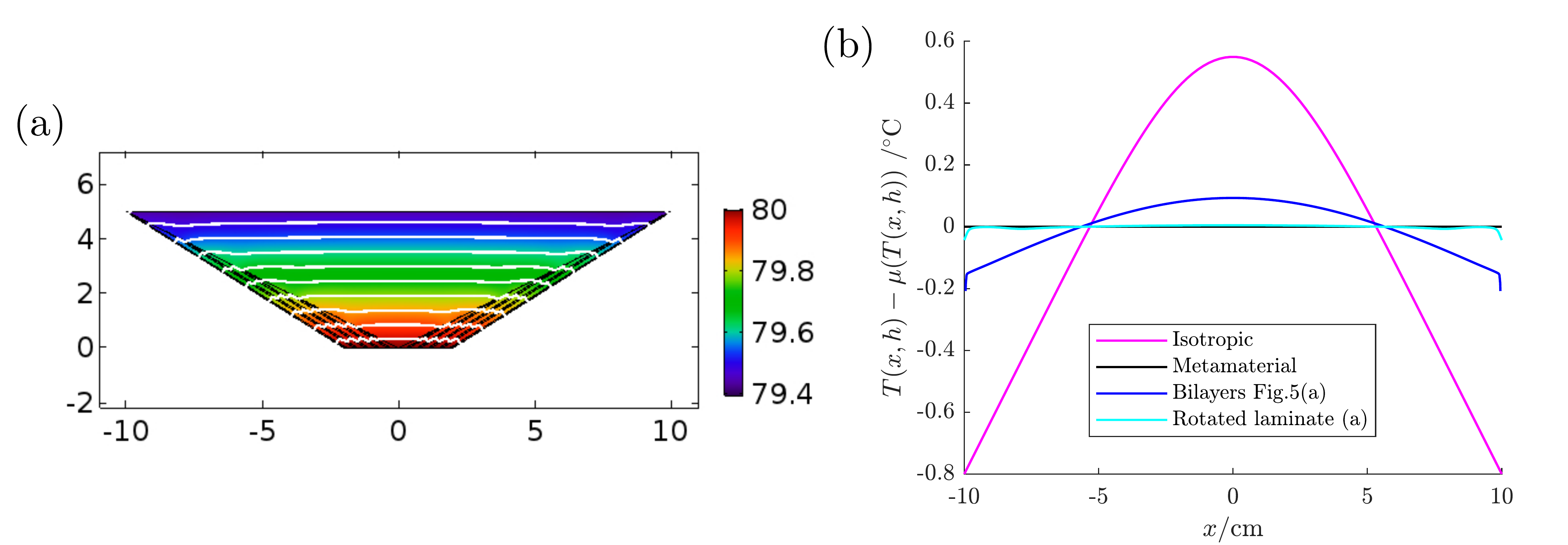}
	\caption{(a) Simulation of a trapezoidal spreader designed using a rotated laminate with 5 bilayers in perfect contact. (b) Comparison of the temperature profile across the top surface for the simulation in (a), the bilayer simulation in Fig.~\ref{fig: 2D results a}(a), and the isotropic and metamaterial simulations from Fig.~\ref{fig: iso vs metamat Cart}.}
	\label{fig: 2D results b}
	\end{center}
\end{figure*}

Fig.~\ref{fig: 2D results b}(a) shows the simulation for the \RED{perfect contact case}. Fig.~\ref{fig: 2D results b}(b) compares the temperature profile across the top surface of the simulations in Fig.~\ref{fig: 2D results a}(a) and Fig.~\ref{fig: 2D results b}(a) against the isotropic and metamaterial simulations from Fig.~\ref{fig: iso vs metamat Cart}. Table~\ref{tab: 2D L2 norms} compares the variation of each temperature profile in Fig.~\ref{fig: 2D results b}(b) with respect to the L2-norm. We see that, when compared to the isotropic case, the temperature variation across the top surface has decreased by approximately $82.1\%$ with the bilayer design and $98.8\%$ with the rotated laminate. In fact, for a sufficiently large number of layers \RED{in perfect contact}, the rotated laminate design will eliminate temperature variation across the top surface since, as we increase the number of layers, this design better represents the required homogeneous conductivity tensor, given by \eqref{eqn: k Cart GC}. On the other hand, the bilayer approximation is unable to eliminate temperature variation completely, even if an infinite number of bilayers are used. This is due to removing $x$-dependence in order to approximate the required inhomogeneous conductivity tensor, given by \eqref{eqn: k Cart linear}.

\begin{table}[h!]
	\caption{Variation of the temperature profiles in Fig.~\ref{fig: 2D design}(f).}
	\centering
	\begin{tabular}{  l  c }
	    \hline
		Design & Temperature Variation /$^{\circ}$C \\
		\hline
        Isotropic & 0.4233\\
        
		Metamaterial & 0 \\
		
		Bilayers & 0.0761 \\
		
		Rotated laminate & 0.0051 \\
		\hline
	\end{tabular}
	\label{tab: 2D L2 norms}
\end{table}

Finally, we recall that the designs implemented for the simulations in Fig.~\ref{fig: 2D results a}(a) and Fig.~\ref{fig: 2D results b}(a) represent the cross section of a spreader in the form of a trapezoidal prism. Therefore, to obtain the full design, we simply extend these configurations in the \text{$y$-direction}. Next we follow analogous steps to design a conical spreader.

\subsection{Conical design}

\paragraph{Linear stretch} 
Here we apply the same approximation as in Section \ref{sec: linear Cart method}, that is, we divide the virtual domain into $n$ sub-layers of equal height and stretch each layer individually where, in this case, the stretch applied to layer $i$ is obtained by substituting   \eqref{eqn: zi} directly into the transformation in   \eqref{eqn: linear map con}. In other words, in layer $i$ we apply the transformation
\begin{equation}
r = \left(\dfrac{(a-b)(2i-1) + 2nb}{2nb} \right)r', \qquad \theta=\theta', \qquad z=z'.
\end{equation}
As a result, the required conductivity in layer $i$ is given by
\begin{equation}
    \boldsymbol{k}_i = k' \left[ \begin{matrix}
    \ 1 & 0 & 0 \ \\
    \ 0 & 1 & 0 \ \\
    \ 0 & 0 & \gamma_i^{-2} \
    \end{matrix}\right],
    \label{eqn: k linear con approx}
\end{equation}
where $\gamma_i$ is given in \eqref{eqn: k linear Cart approx}.

\begin{figure*}[p!]
	\begin{center}
	\includegraphics[width=.9\textwidth]{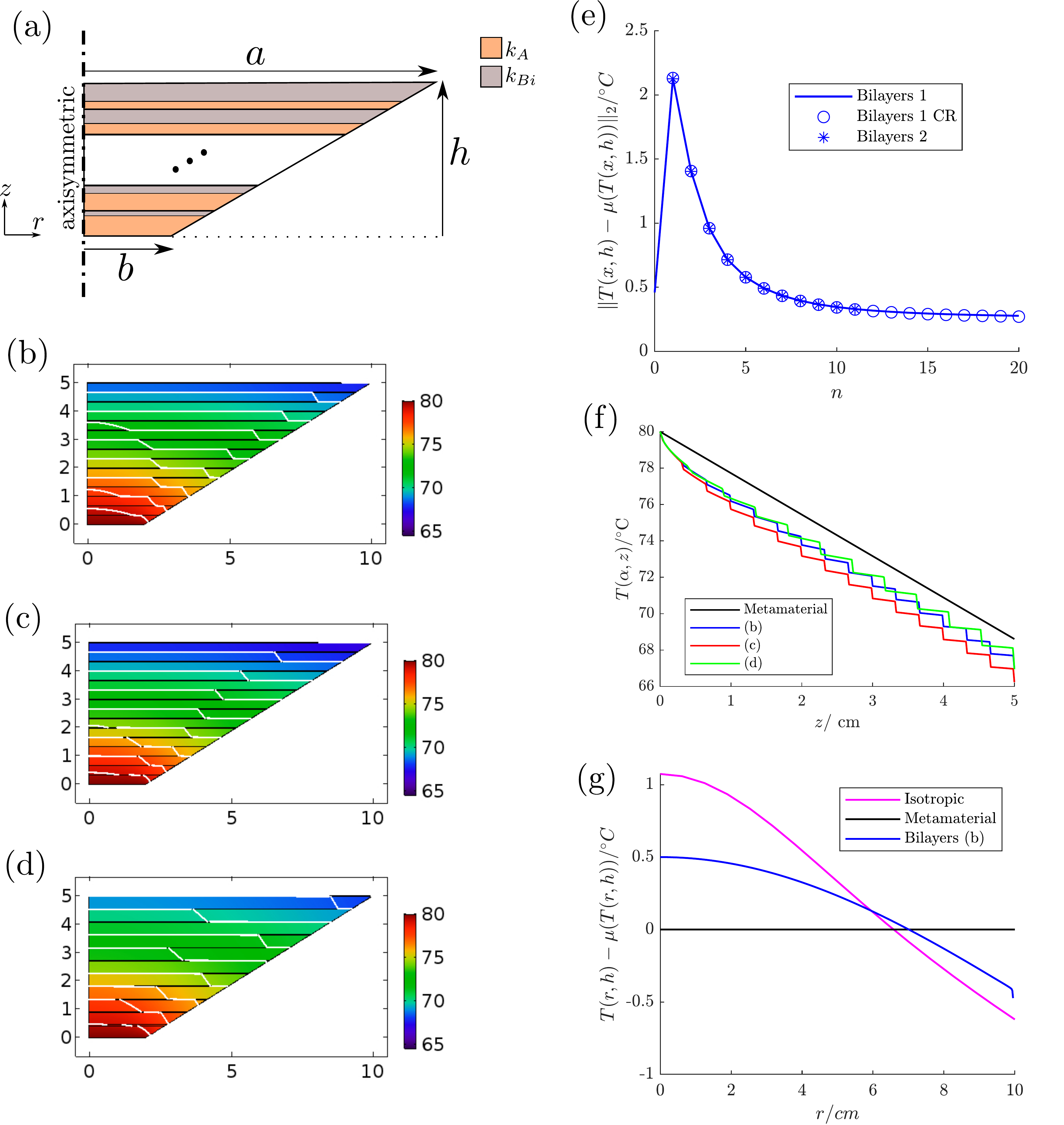}
	\caption{\RED{Designs and simulations to analyse conical spreaders: (a) Conical heat spreader design using bilayers; (b)-(c) simulations of a conical spreader designed using $n=15$ bilayers in perfect contact, however (c) has $R_c$ added to interfaces; (d) simulation of a conical spreader with $R_c$ incorporated into the design; (e) simulations comparing the temperature variation across the top surface using $n$ bilayers \RED{for three cases: a case designed for perfect contact; the same perfect contact design, but with thermal contact resistance between each layer; and a second design where $R_c$ is incorporated into $k_{Bi}$.}; (f) comparison of the temperature drop along the external surfaces ($T(\alpha,z)$ where $\alpha = z(a-b)/h +b$) of the simulations in (b)-(d) against the metamaterial in Fig.~\ref{fig: iso vs metamat con}(b); (g) comparison of the temperature profile across the top surface of (b) with the isotropic and metamaterial simulations from Fig.~\ref{fig: iso vs metamat con}.}}
	\label{fig: 3D approx}
	\end{center}
\end{figure*}

The anisotropic, homogeneous behaviour in \eqref{eqn: k linear con approx} can be approximated with a bilayered material by substituting $k_\parallel=k'$ and $k_\perp=k'/\gamma_i^2$ into \eqref{eqn: eff k}. This design is illustrated in Fig.~\ref{fig: 3D approx}(a) where the layers lie parallel to the $r$-axis. As in Section \ref{sec: linear Cart method}, for this design the conductivity of each bottom layer, denoted by $k_A$, is fixed throughout whereas the conductivity of each top layer, denoted by $k_{Bi}$ for layer $i$, varies from bilayer to bilayer and is calculated from \eqref{eqn: kB}. The necessary filling fraction in each bilayer is then calculated from \eqref{eqn: phi}. For a given configuration we then run simulations for a range of $n$ bilayers and calculate the temperature variation across the top surface with respect to the L2-norm, taking into account the axisymmetric geometry for this case. 

\RED{For proof of concept, this process is shown in Fig.~\ref{fig: 3D approx}(e) for the same three cases as in Section~\ref{sec: linear stretch approx}} when $a=5b=2h=10$cm$,\ k'=80$W/(mK)$,\ k_A=1.05k'$ and the same boundary conditions as in Fig.~\ref{fig: iso vs metamat con} are imposed. \RED{Note that, when $n>11$, \eqref{eqn: L bound} is not satisfied for the second bilayer design where $R_c$ is built in. Therefore, there are no results for these cases. Once again, we see that the effects of thermal contact resistance are negligible in terms of the temperature variation. From Fig.~\ref{fig: 3D approx}(e) we can choose a suitable choice for $n$, for example, a choice of $n=15$ for the first two cases and $n=11$ for the case where $R_c$ is built in achieves a temperature variation of approximately $0.29^\circ$C. Fig.~\ref{fig: 3D approx}(b)-(d) shows simulations for these cases respectively. The parameters for each bilayer in Fig.~\ref{fig: 3D approx}(b)-(c) and Fig.~\ref{fig: 3D approx}(d) are provided in columns 3-4 and 6-7 of Table \ref{tab: 3D bilayers} respectively.} Note that, $k_\parallel$ and $k_A$ are omitted from Table \ref{tab: 3D bilayers} as they are the same in each bilayer, namely $k_\parallel=k'=80$W/(mK) and $ k_A=1.05k'=84$W/(mK).

\begin{table}[h!]
	\caption{Parameters for the bilayer simulations in Fig.~\ref{fig: 3D approx}(c).}
	\centering
\begin{tabular}{c c c c c c c}
	\hline
Bilayer &  $k_\perp \ /(W/(mK))$  &  $k_B \ /(W/(mK))$  & \hspace{.75cm}$\phi$\hspace{.75cm} &  \RED{$k_\perp^{CR} \ /(W/(mK))$}  & \RED{$k_B^{CR} \ /(W/(mK))$} & \hspace{.75cm} \RED{$\phi^{CR}$} \hspace{.75cm} \\
		 \hline
1 & 62.28  &	11.47 &   0.95  & \RED{57.28}	& \RED{41.27} &	\RED{0.91} \\
2 & 40.82  &	3.78  &	  0.95  & \RED{33.49}	& \RED{3.51}  &	\RED{0.95} \\
3 & 28.80   &	2.09  &	  0.95  & \RED{21.95}	& \RED{1.63}  &	\RED{0.95} \\
4 & 21.400   &	1.37  &	  0.95  & \RED{15.49}	& \RED{0.99}  &	\RED{0.95} \\
5 & 16.53  &	0.98  &   0.95  & \RED{11.51}	& \RED{0.67}  &	\RED{0.95} \\
6 & 13.15  &	0.74  &   0.95  & \RED{8.89}	& \RED{0.50}	&   \RED{0.95} \\
7 & 10.71  &	0.58  &	  0.95  & \RED{7.07}	& \RED{0.38}  &	\RED{0.95} \\
8 & 8.89   &	0.47  &	  0.95  & \RED{5.76}	& \RED{0.30}	&   \RED{0.95} \\
9 & 7.50	   &    0.39  &	  0.95  & \RED{4.78}	& \RED{0.25}  &	\RED{0.95} \\
10 & 6.41   &	0.33  &	  0.95  & \RED{4.03}	& \RED{0.21}  &	\RED{0.95} \\
11 & 5.54   &	0.28  &	  0.95  & \RED{3.45}	& \RED{0.17}  &	\RED{0.95} \\
12 & 4.84   &	0.21  &	  0.95  & - & - & - \\
13 & 4.26   &	0.21  &   0.95  & - & - & - \\
14 & 3.78   &	0.19  &	  0.95  & - & - & - \\
15 & 3.38   &	0.17  &	  0.95  & - & - & - \\
\hline
\end{tabular}
	\label{tab: 3D bilayers}
\end{table}

\RED{Fig.~\ref{fig: 3D approx}(e) compares the temperature drop along the external surface of the simulations in Fig.~\ref{fig: 3D approx}(b)-(d) against the metamaterial in Fig.~\ref{fig: iso vs metamat con}(b). We see that when the effects of contact resistance are introduced to the perfect contact design a larger temperature drop occurs. By incorporating the effects of contact resistance into our model we have been able to better approximate the true metamaterial design. As a result, we can account for part of the reduction in temperature when experiments are compared to simulations in \cite{Liu2014}.}

Fig.~\ref{fig: 3D approx}(g) compares the temperature profile across the top surface of Fig.~\ref{fig: 3D approx}(b) against the isotropic and metamaterial simulations from Fig.~\ref{fig: iso vs metamat con}. Table~\ref{tab: 3D L2 norms} compares the variation of each temperature profile in Fig.~\ref{fig: 3D approx}(g) with respect to the L2-norm, taking into account the axisymmetric geometry. We see that, when compared to the isotropic case, the temperature variation across the top surface has decreased by approximately $35.8\%$ with the bilayer design \RED{in perfect contact. It is evident that the $r$-dependence in the conical metamaterial is more influential than the $x$-dependence in the trapezoidal metamaterial.} Since we remove the $r$-dependence to approximate the required conductivity tensor, given by \eqref{eqn: k con linear}, we are unable to eliminate temperature variation completely, even if an infinite number of layers \RED{in perfect contact} are used.

\begin{table}[h!]
	\caption{Variation of the temperature profiles in Fig.~\ref{fig: 3D approx}(g).}
	\centering
	\begin{tabular}{  l  c }
	    \hline
		Design & Temperature Variation /$^{\circ}$C \\
		\hline
        Isotropic & 0.4597\\

		Metamaterial & 0 \\

		Bilayers & 0.2952 \\
		\hline
	\end{tabular}
	\label{tab: 3D L2 norms}
\end{table}


The designs in this section better represent each metamaterial (or approximation of a metamaterial) as we increase the number of layers, however, as we increase the number of layers we also increase the number of interfaces - and therefore the effects of thermal contact resistance. This motivates the next investigation where we aim to achieve spatially uniform thermal fields whilst minimising the number of interfaces. We do this by simply splitting the spreader into two or three isotropic, homogeneous components and exploiting the solution to the diffusion equation in this geometry. The results of which are tailored to a specific set of boundary conditions.

\section{Neutral layer method and validation}
\begin{figure}[h!]
	\begin{center}
	\includegraphics[width=0.95\textwidth]{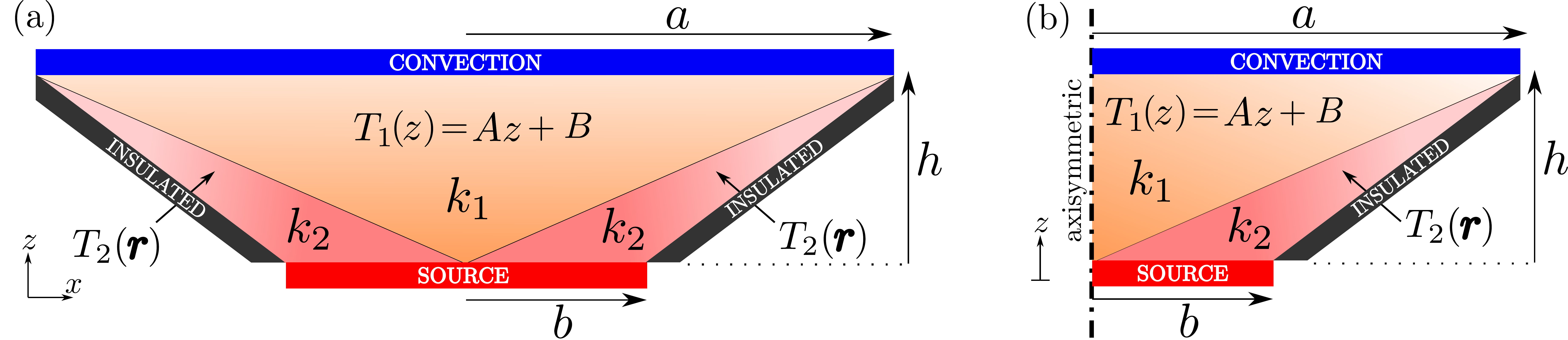}
	\caption{Two-dimensional cross section of a neutral layer design for: (a) trapezoidal heat spreader; (b) conical heat spreader. Each component has an isotropic, homogeneous conductivity denoted by a constant scalar $k_1$ or $k_2$.}
	\label{fig: NL concepts}
	\end{center}
\end{figure}

Here we propose a conical and trapezoidal heat spreader design composed of two or three components respectively. Furthermore, each component has an isotropic, homogeneous conductivity, denoted by the constant scalars $k_1$ or $k_2$. The cross section of these designs are illustrated in Fig.~\ref{fig: NL concepts}, where Fig.~\ref{fig: NL concepts}(a) is extended in the $y$-direction to obtain the trapezoidal design and Fig.~\ref{fig: NL concepts}(b) is rotated about the $z$-axis to obtain the conical design. We assume that the temperature fields in the upper (orange) domains, denoted by $T_1$, have the desired form, however, we drop this assumption for the temperature fields in the lower (pink) domains, denoted by $T_2$. By allowing $T_2$ to depend on $x$ (or $r$) we are able to solve the inverse problem where we achieve the desired result in the upper components by engineering the ratio between $k_1$ and $k_2$. When $k_1$ and $k_2$ complement each other appropriately, the lower layers neutralise any perturbations caused by the funnel-shaped geometry, hence we refer to them as neutral layers. We first consider the trapezoidal design.

\subsection{Trapezoidal design}
\begin{figure}[t!]
	\begin{center}
	\includegraphics[width=0.9\textwidth]{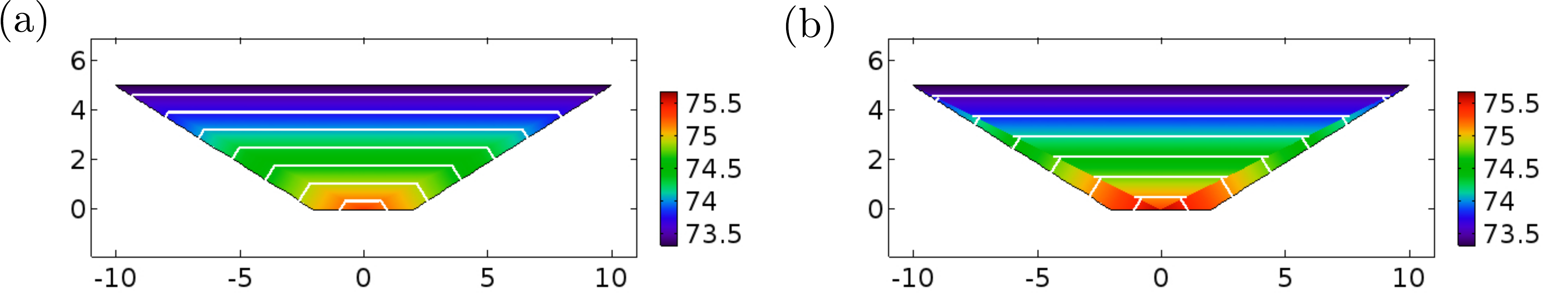}
	\caption{Simulations of a trapezoidal heat spreader designed using neutral layers with: (a) Perfect contact\RED{; and (b) thermal contact resistance when $R_c=4\times10^{-4}$m$^2$K/W.}}
	\label{fig: heat map NL Cart}
	\end{center}
\end{figure}

The trapezoidal design is composed of three isotropic, homogeneous components. For this case, when a constant heat flux, given by $q_0$, is imposed across the base, an analytic solution can be found \RED{with and without thermal contact resistance} where
\RED{\begin{equation}
    T_1(z)=-\dfrac{b}{a}\dfrac{q_0}{k_1}z + \dfrac{b}{a}\dfrac{q_0}{k_1}\left(\dfrac{k_1}{h_c} + h\right) + T_0,
    \label{eqn: T1 NL Cart}
\end{equation}}
on the condition that
\begin{equation}
    \dfrac{k_2}{k_1} = \dfrac{a(h^2 + a(a-b))}{bh^2},
    \label{eqn: k ratio}
\end{equation}
where $a,\ b$ and $h$ are shown in Fig.~\ref{fig: NL concepts}(a). Details of these results are provided in \ref{app: NL cart} along with the resulting temperature field for $T_2(x,z)$. \RED{In addition, the refraction law of heat flux, employed in \cite{Hu2015} to design an anisotropic ground cloak, can be used to verify the result in \eqref{eqn: k ratio}.

From \eqref{eqn: T1 NL Cart} and \eqref{eqn: k ratio} we can see that, for this set of boundary conditions, $T_1$ and $k_2/k_1$ are unaffected by thermal contact resistance and, as a result, the uniform temperature across the top surface is unaffected. Therefore, the temperature drop caused by thermal contact resistance solely affects $T_2$. This is evident from \eqref{A3 E} where we can see that $T_2$ depends linearly on $R_c$ when all other parameters are fixed. For example, Fig.~\ref{fig: heat map NL Cart}(a) and (b) show simulations for a trapezoidal spreader designed using neutral layers in perfect contact and with contact resistance respectively.} For both simulations we set $a=5b=2h=10$cm and impose a constant heat flux across the base, given by $q_0 = 4,000$W/m$^2$, and a convective boundary condition across the top surface where the surrounding air has heat transfer coefficient $h_c = 15$W/(m$^2$K) and temperature $T_0=20^{\circ}$C. Using \eqref{eqn: k ratio}, we choose $k_1=20$W/(mK) and set $k_2 = 21k_1$.
We see that the thermal fields in the upper component of both simulations are spatially uniform with respect to $x$, eliminating any temperature variation across the top surface, as desired. \RED{Note that, for the case with contact resistance, we set $R_c=4\times10^{-4}$m$^2$K/W to better visualise the temperature jump, which is given by $[\![ T ]\!]_{\mathbf{r}_I} = bq_0 R_c/\sqrt{h^2+a^2} \approx 716$W/m$^2 \times R_c$ for this configuration, as discussed in \ref{app: NL cart}. Therefore, the effects of contact resistance are negligible for $R_c<4 \times 10^{-4}$m$^2$K/W since $[\![ T ]\!]_{\mathbf{r}_I} < 0.29$K. This is to be expected since the design only has one interface. As a result, we assume that $R_c$ does not affect either the Cartesian or conical neutral layer designs at this scale.}

Note that, further analytical solutions of the form $T_1=Az+B$ cannot be found for the trapezoidal or conical neutral layer design when either a constant heat flux or constant temperature is imposed across the base, however, we are able to optimise a design for a given set of parameters. For example, next we demonstrate how to optimise the neutral layer design for a conical spreader when a constant temperature is imposed across the base. We consider the optimal design to be the one that minimises the temperature variation across the top surface with respect to the L2-norm. The methods used here are also applicable to the trapezoidal design when a constant temperature is forced across the base.

\subsection{Optimal conical design}
We can approach the conical design process in one of two ways: choose the two materials and find the optimal geometry, or preset the geometry and find the optimal ratio between $k_1$ and $k_2$. Here we choose the latter and preset the geometry. Furthermore, we specify our choice of $k_1$ and the parameters for each boundary condition such that $k_2$ is the only unknown parameter. To find the optimal ratio between the two conductivities we run simulations for a range of $k_2$ and calculate the temperature variation across the top surface each time with respect to the L2-norm, taking into account the axisymmetric geometry of the spreader. Once complete, we simply find the ratio that achieves the minimum temperature variation and declare this to be the optimal configuration.

\begin{figure}[t!]
	\begin{center}
	\includegraphics[width=0.9\textwidth]{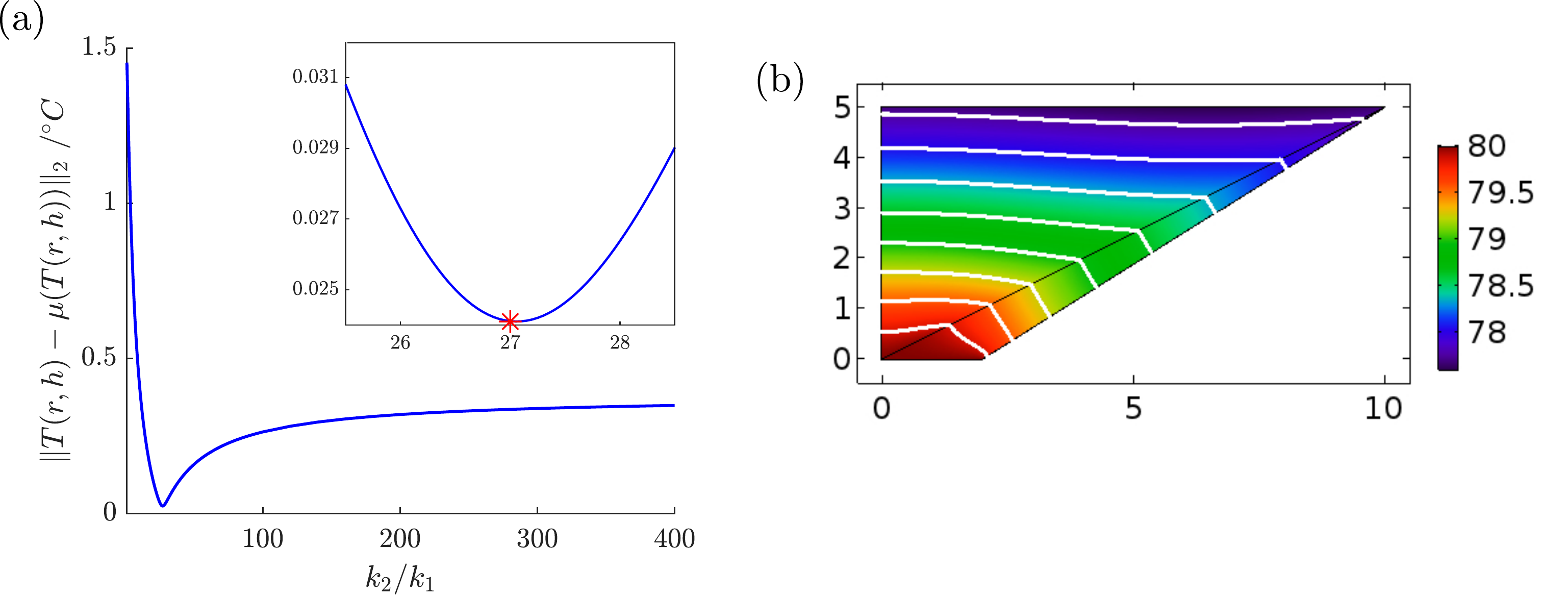}
	\caption{\RED{(a) Simulations to determine the optimal ratio between $k_1$ and $k_2$ for a conical neutral layer design. Inset showing results near the global minimum.} (b) Simulation for optimal case when $k_2=27k_1$.}
	\label{fig: heat map NL con}
	\end{center}
\end{figure}

This process is shown in Fig.~\ref{fig: heat map NL con}(a) where we set $a=5b=2h=10$cm and $k_1=20$W/(mK). For these simulations we impose a constant temperature across the base, given by $T(z=0)=80^{\circ}$C, and the same convective condition as in Fig.~\ref{fig: heat map NL Cart}. We find that, for this configuration, the optimal ratio between the conductivities occurs when $k_2\approx 27k_1$. \RED{Note that, the optimal ratio depends heavily on the geometry, not on our choice of $k_1$. For example, when $k_1>0.5$W/(mK), $26.5<k_2/k_1 < 27$ for this configuration.} Fig.~\ref{fig: heat map NL con}(b) shows a simulation of the resulting thermal fields through the cross section of the spreader when $k_2=27k_1$. The temperature variation across the top surface for this optimal case is $0.0241 ^{\circ}$C. Therefore, by incorporating a neutral layer with isotropic conductivity $k_2=27k_1$, we have reduced the temperature variation by over $98\%$ when compared to the isotropic case ($k_2/k_1=1$ in Fig.~\ref{fig: heat map NL con}(a)) where the temperature variation is approaching $1.5 ^{\circ}$C. This being said, we are yet to completely eliminate the temperature variation across the top surface. In order to optimise further, next we demonstrate how to incorporate curvature into our design.

\begin{figure}[b!]
	\begin{center}
	\includegraphics[width=.8\textwidth]{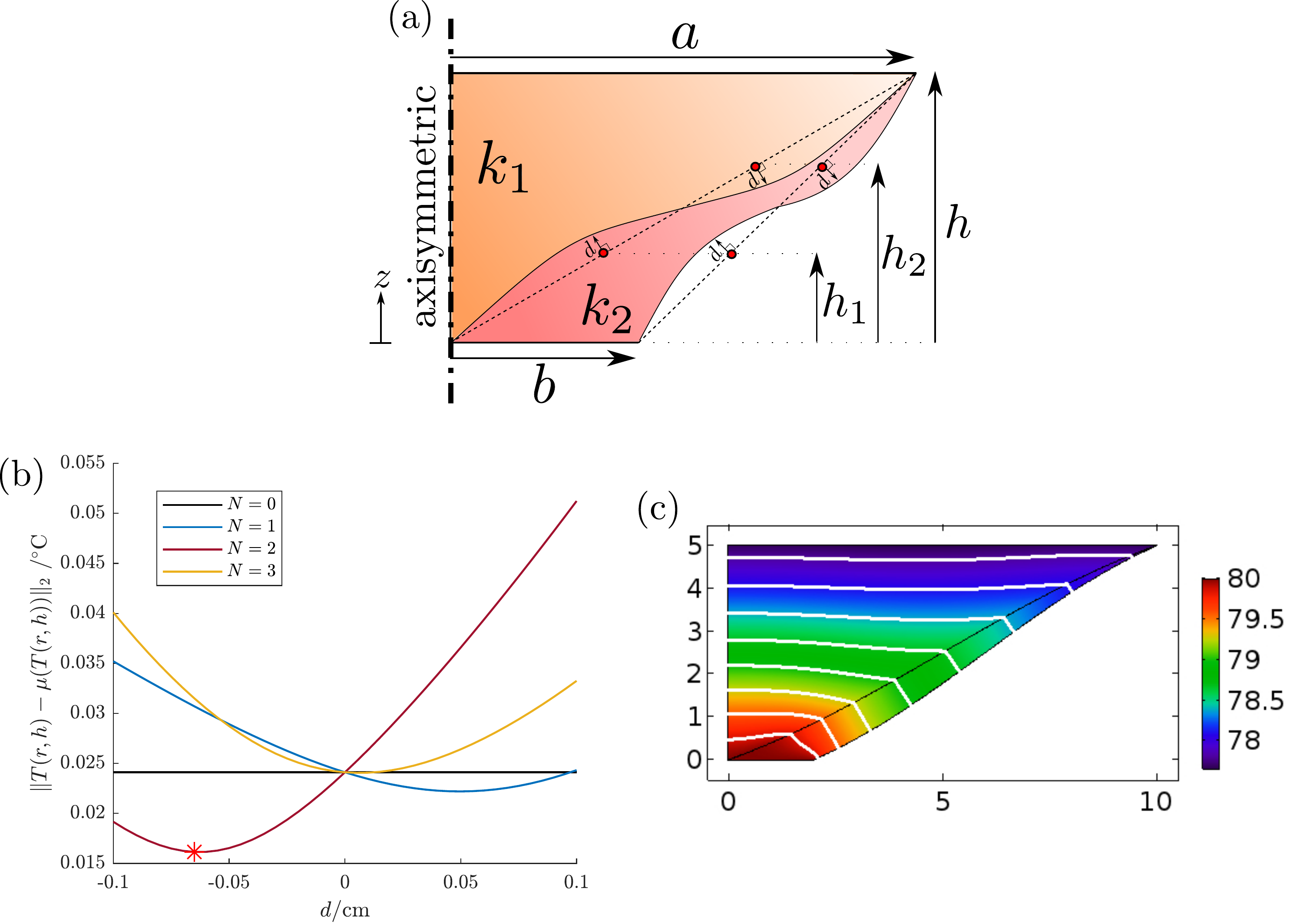}
	\caption{(a) Conical heat spreader with curved neutral layer design. (b) Simulations to determine the optimal number of points, $N$, to add curvature, $d$, to. (c) Simulation for optimal case when $N=2$ and $d=-0.065$cm.}
	\label{fig: curvature concept}
	\end{center}
\end{figure}

\subsubsection{Incorporating curvature}
We can add curvature to one interface at a time or both interfaces simultaneously. Here we incorporate curvature into both interfaces by selecting $N$ equally spaced points between 0 and $h$ and altering their position by a distance $d$ in a direction perpendicular to the interface. This concept is illustrated in Fig.~\ref{fig: curvature concept}(a) for when $N=2$. 

To find the optimal curvature for a given configuration we fix all the parameters apart from $N$ and $d$. We then run simulations for different values of $N$ and a range of $d$, calculating the temperature variation across the top surface each time in order to find the combination that achieves the minimal temperature variation.
This process is shown in Fig.~\ref{fig: curvature concept}(b) where we set the same geometry and boundary conditions as in Fig.~\ref{fig: heat map NL con}(b). Here we set $N=0,1,2,3$ and then run simulations for a range of $d$ where $-0.1$cm~$\leq d \leq$~$0.1$cm. We see that, for this configuration, the minimum temperature variation of $0.0161^{\circ}$C occurs when  $N=2$ and $d=-0.065$cm. Fig.~\ref{fig: curvature concept}(c) shows a simulation of the geometry and resulting thermal fields for this optimal case. By adding curvature in this way we have been able to reduce the temperature variation by a further $33.2 \%$ when compared to the simulation in Fig.~\ref{fig: heat map NL con}(b) with no curvature.

\section{Conclusion}
We have described a class of potentially realisable heat spreaders composed from isotropic, homogeneous materials. Each spreader has been designed to ensure that thermal fields are uniform across a large convective surface in a steady-state regime when forced by a smaller thermal source. Two categories of heat spreaders have been proposed: the first using the concept of transformation thermodynamics; and the second using the optimisation of a simple inverse problem, a configuration referred to as a neutral layer. All designs have been validated numerically \RED{and the effects of thermal contact resistance have been discussed on a case-by-case basis.

The transformed spreaders that are designed as a result of a linear stretch are the most affected by contact resistance. This is due to the many interfaces lying perpendicular to the direction of the heat flux. By extending previous models to include contact resistance we have shown that, whilst the effects of contact resistance are negligible in terms of temperature variation, we can account for part of the reduction in temperature when experiments are compared to simulations in \cite{Liu2014}. As a result, our models help to improve the approximation of the metamaterial designs. Furthermore, by simply modifying a ground cloak transformation we have been able to utilise the mapping in a new heat-spreading context. By incorporating contact resistance into our model we have been able to show that the transformed spreaders that are designed with a modified ground cloak are less affected by contact resistance. This is to be expected since the rotated laminate design leads to interfaces that are not perpendicular to the direction of the heat flux.

In addition to the transformed designs, our neutral layer designs help gain insight into the unique arrangement of natural materials that can be employed to guide the flow of heat through a funnel-shaped design in a uniform manner. The motivation behind these simple configurations was to minimise the number of interfaces and therefore minimise the effects of contact resistance. By adding contact resistance to our model we have been able to verify that, indeed, contact resistance has little effect on these designs, as desired. Furthermore, we have shown that by adding curvature to our neutral layer designs we are able to optimise even further.} These insights will hopefully help broaden research on how to control the flow of heat and help pave the way for future thermal management designs.

In future work we intend to refine the proposed designs by incorporating time-dependence into our mathematical models. Indeed transient effects play an important role in thermal engineering problems, affecting heat loss from buildings \cite{Parnell2016} or influencing thermal convective processes \cite{Gorbushin2019}.

\section{Declaration of Competing Interests}
The authors declare that they have no competing financial and personal relationships that could inappropriately influence the work in this paper.

\section{Acknowledgements}
The authors would like to thank The Department of Mathematics at The University of Manchester for funding Russell's PhD. Parnell is grateful to the Engineering and Physical Sciences Research Council (EPSRC), UK, for funding his Fellowship extension EP/S019804/1.

\appendix
\section{} \label{app: princ sys}
Determining the principal conductivities and principal axes of a conductivity tensor is equivalent to finding its eigenvalues and corresponding eigenvectors. In other words, for a conductivity tensor $\boldsymbol{k}$, where
\begin{equation}
\boldsymbol{k} = \left[ \begin{matrix}
k_{11}& k_{12} \\
k_{12}& k_{22} \\
\end{matrix} \right]
\label{eqn: gen k ani homo}
\end{equation}
for constant scalars $k_{11},\ k_{12},\ k_{12}$ and $k_{22}$, the principal conductivities are the roots of the polynomial $f(\lambda) = \det\left(\boldsymbol{k} - \lambda \mathbf{I}\right)$. For this example we obtain two roots, $\lambda_+$ and $\lambda_-$, given by
\begin{equation}
\lambda_\pm = \dfrac{k_{11} + k_{22} \pm \sqrt{(k_{11} + k_{22})^2 - 4\left(k_{11}k_{22} - k_{12}^2\right)}}{2}.
\label{eqn: k princ}
\end{equation}
In this work, the principal conductivities correspond to $k_\parallel$ and $k_\perp$. Therefore, since $k_\parallel > k_\perp$, we let $k_\parallel=\lambda_+$ and $k_\perp = \lambda_-$. The corresponding principal axes \text{$\mathbf{u} = [u_1,u_2]^T$} and $\mathbf{v} = [v_1,v_2]^T$, illustrated in Fig.~\ref{fig: 2D design}(b), have unit length and satisfy $\left(\mathbf{k} - k_+\mathbf{I}\right)\mathbf{u} = \mathbf{0}$ and $\left(\mathbf{k} - k_-\mathbf{I}\right)\mathbf{v} = \mathbf{0}$ respectively. In particular, $\mathbf{u}$ is given by
\begin{equation}
\label{eqn prin ax}
\mathbf{u} = \left[\sqrt{\dfrac{k_\parallel - k_{22}}{2k_\parallel - k_{22} - k_{11}}} \ , \quad \sqrt{\dfrac{k_\parallel - k_{11}}{2k_\parallel - k_{22} - k_{11}}} \ \right]^T.
\end{equation}
Therefore, since the $\mathbf{u}$ has unit length, we can determine the corresponding rotation by calculating \text{$\cos \theta = u_2/1=u_2$} such that
\begin{equation}
\label{eqn: theta}
\cos \theta = \sqrt{\dfrac{k_\parallel - k_{22}}{2k_\parallel - k_{22} - k_{11}}}.
\end{equation}
When the conductivity tensor is aligned with its principal axes it can be written in the form $\boldsymbol{k}=\textnormal{diag}(k_\parallel,k_\perp)$.

\section{}\label{app: NL cart}
We look for a solution to the configuration in Fig. \ref{fig: NL concepts}(a) with the form $T_1(z) = Az + B$ and $T_{2}(x,z) = Cx +Dz+E$ where $A,\ B,\ C,\ D$ and $E$ are constants. Note that, since $k_1$ and $k_2$ are constant scalars for this example, the heat diffusion equation reduces to Laplace's equation for which $T_1$ and $T_2$ are known solutions. Therefore, we are left to satisfy the following boundary conditions:

Firstly, the insulated external surfaces require $\nabla T_2 \cdot \mathbf{n}_1 = 0$ where 
\begin{equation}
\textbf{n}_1 = \left(\left(\frac{h}{a-b}\right)^2 +1 \right)^{-\frac{1}{2}} \left[\textnormal{sgn}(x) \dfrac{h}{a-b},\ -1 \right]^T,
\end{equation}
is the outward pointing unit normal. Therefore,
\begin{equation}
\left(\textnormal{sgn}(x)\dfrac{h}{a-b} C - D \right) = 0 \qquad \implies \qquad C = \textnormal{sgn}(x)\dfrac{(a-b)}{h}D.
\label{eqn: C D}
\end{equation}
\RED{Next we incorporate thermal contact resistance into the model by assuming that a temperature jump occurs across each interface. Let $\mathbf{r}_I$ respresent all the points where $z=h|x|/a=\textnormal{sgn}(x)hx/a$. Then the temperature jump is given by
\begin{equation}
    [\![ T ]\!]_{\mathbf{r}_I} = T(\mathbf{r}_I^-) - T(\mathbf{r}_I^+) = R_c \left( -k \nabla T \cdot \mathbf{n}_2 \right) \rvert _{z=\mathbf{r}_I^+},
    \label{app: temp jump}
\end{equation}
where $R_c$ is the thermal resistance of the interface and $\mathbf{n}_2$ is the unit normal given by
\begin{equation}
\textbf{n}_{2} = \left(\left(\frac{h}{a}\right)^2 +1 \right)^{-\frac{1}{2}} \left[ -\textnormal{sgn}(x) \dfrac{h}{a},\ 1 \right]^T.
\end{equation}}
Therefore, we require
\begin{equation}
\textnormal{sgn}(x)\dfrac{a}{h}Cz + Dz + E = Az + B \RED{- R_c k_1 A\left(\left(\frac{h}{a}\right)^2 +1 \right)^{-\frac{1}{2}}} = Az + B \RED{- \dfrac{a R_c k_1 A}{\sqrt{h^2 +a^2}}}.
\end{equation}
Then, using \eqref{eqn: C D}, we obtain
\begin{equation}
\left(\dfrac{a}{h}\dfrac{(a-b)}{h} + 1 \right) Dz +  E = Az + B \RED{- \dfrac{a R_c k_1 A}{\sqrt{h^2 +a^2}}},
\end{equation}
such that
\begin{equation}
A = \left(\dfrac{ (a-b)a }{h^2} + 1 \right) D= \left(\dfrac{ (a-b)a +h^2 }{h^2} \right)D \qquad \text{and} \qquad E = B \RED{- \dfrac{a R_c k_1 A}{\sqrt{h^2 +a^2}}}.
\label{A4 a}
\end{equation} 
Now we impose continuity of flux across the same interfaces such that $k_1 \nabla T_1 \cdot \mathbf{n}_2 = k_2 \nabla T_2 \cdot \mathbf{n}_2$.
As a result, we require
\begin{equation}
\quad k_1A = k_2 \left( -\textnormal{sgn}(x)\dfrac{h}{a} C + D \right) = k_2 \left(- \dfrac{(a-b)}{a} +1 \right) D = k_2 \dfrac{b}{a} D.
\end{equation}
Therefore, after substitution of $A$ from \eqref{A4 a} and some rearranging, we have the condition that
\begin{equation}
k_2 =\left(\dfrac{ a(h^2+ a(a-b)) }{bh^2} \right) k_1.
\label{eqn: k ratio app}
\end{equation}

Next we impose a constant heat flux, $q_0$W/m$^2$,  across the base such that $q_0 = -k_2 \nabla T_2 \cdot \mathbf{e}_z$ when $z=0$. In other words,
\begin{equation}
 q_0 = -k_2 D \qquad \implies \qquad D = -\dfrac{q_0}{k_2}.
\end{equation}
Finally, we impose a convective condition across the top surface such that $-k_1 \nabla T_1 \cdot \mathbf{e}_z = h_c \left[T_1 - T_0 \right]$ when $z=h$ where $h_c$ and $T_0$ are the heat transfer coefficient and temperature of the convective fluid respectively. As a result
\begin{equation}
-k_1A = h_c \left[ Ah + B -T_0 \right] \qquad
 \implies \qquad -A \left(\dfrac{k_1}{h_c} + h \right) + T_0 = B.
\end{equation}
Therefore, provided the ratio between $k_1$ and $k_2$ satisfies \eqref{eqn: k ratio app}, we can find a solution of the form $T_1(z)=Az+B$ and $T_2(x,z)=C|x|+Dz+E$ where the constants $A,\ B,\ C,\ D$ and $E$ are given by

\begin{equation}
    D = -\dfrac{q_0}{k_2},
\end{equation}
\begin{equation}
A = \left( \dfrac{h^2 + a(a-b)}{h^2} \right) D = \dfrac{b}{a}\dfrac{k_2}{k_1} D = -\dfrac{b}{a}\dfrac{q_0}{k_1},
\label{A3 A}
\end{equation}

\begin{equation}
C = \textnormal{sgn}(x) \dfrac{(a-b)}{h}D = - \textnormal{sgn}(x) \dfrac{(a-b)}{h}\dfrac{q_0}{k_2},
\end{equation}

\begin{equation}
B = -A \left(\dfrac{k_1}{h_c} + h \right) + T_0 = \dfrac{b}{a}\dfrac{q_0}{k_1}\left(\dfrac{k_1}{h_c} + h \right) + T_0.
\label{A3 B}
\end{equation}

\begin{equation}
E = B \RED{- \dfrac{a R_c k_1 A}{\sqrt{h^2 +a^2}}} = \dfrac{b}{a}\dfrac{q_0}{k_1}\left(\dfrac{k_1}{h_c} + h \right) + T_0  \RED{+ \dfrac{b R_c q_0}{ \sqrt{h^2 + a^2}}}.
\label{A3 E}
\end{equation}

\bibliography{References}

\end{document}